# On an Improvement over Rényi's Equivocation Bound

**Nandakishore Santhi**
Department of Electrical and Computer Engineering
University of California San Diego
9500 Gilman Drive, La Jolla, CA 92093
nsanthi@ucsd.edu

**Alexander Vardy**
Department of Electrical and Computer Engineering
University of California San Diego
9500 Gilman Drive, La Jolla, CA 92093
vardy@kilimanjaro.ucsd.edu

*Abstract*— We consider the problem of estimating the probability of error in multi-hypothesis testing when MAP criterion is used. This probability, which is also known as the Bayes risk is an important measure in many communication and information theory problems. In general, the exact Bayes risk can be difficult to obtain. Many upper and lower bounds are known in literature. One such upper bound is the equivocation bound due to Rényi which is of great philosophical interest because it connects the Bayes risk to conditional entropy. Here we give a simple derivation for an improved equivocation bound.

We then give some typical examples of problems where these bounds can be of use. We first consider a binary hypothesis testing problem for which the exact Bayes risk is difficult to derive. In such problems bounds are of interest. Furthermore using the bounds on Bayes risk derived in the paper and a random coding argument, we prove a lower bound on equivocation valid for most random codes over memoryless channels.

## I. INTRODUCTION

In his celebrated paper of 1948, Shannon proved the Channel Coding Theorem. This theorem essentially states that the ensemble of long random block codes (and thus *some* specific code) in the limit of very large block lengths, achieves an arbitrarily low probability of error under decoding by jointly typical decision rule, when used over a given channel at information rates below a limit called the channel's Shannon capacity. It is well known that for minimizing the Bayes risk, the optimal decision rule is the *Maximum Áposteriori Probability (MAP)* decision rule. Shannon uses jointly typical decision rule in his analysis because, asymptotically the decision rule is optimal and it simplifies the analysis considerably. The strong converse to the channel coding theorem based on Fano's inequality states that the probability of error under *any* decision rule approaches 1 exponentially as block length increases when rate is above capacity.

The Shannon capacity of a discrete memoryless channel (DMC) is given by,

$$C = \max_{p_x(.)} I(X;Y)$$

where $I(X;Y)$ is the mutual information between the channel input $X$ and channel output $Y$. The mutual information is given in terms of entropy function as,

$$\begin{aligned} I(X;Y) &= H(X) - H(X|Y) \\ &= \sum_{x,y} p(x,y) \log \frac{p(x,y)}{p(x)p(y)} \end{aligned} \quad (1)$$

The source entropy $H(X)$ is a function of the source statistics. The function $H(X|Y)$ is called the conditional entropy or equivocation. Equivocation is dependent on the channel statistics as well as the properties of the channel code employed. For most non-trivial channels, computation of capacity is infeasible due to the optimization required over the input probability distribution of a highly nonlinear function. Good upper and lower bounds to capacity which are easy to compute are therefore of interest. A useful lower-bound on capacity is clearly the mutual information for some arbitrary $p(x)$. Upper bounds usually require other formulations.

The decoding problem for codes is an instance of the more general problem of multiple hypothesis testing which appears in some form in most fields of science. It is intuitive to say that the probability of error under the Bayes decision rule is a function of equivocation. That this is true was proved rigorously by Rényi in [R66]. Among other things, he showed that $P_e \leqslant H(X|Y)$. Hellman and Raviv later improved on this result in [HR70] and showed that in fact, $P_e \leqslant \frac{1}{2} H(X|Y)$. It is immediately clear that even this improved bound is extremely loose when the equivocation is over unity.

In this paper we first look at several tight classical bounds on the Bayes risk in the general multi-hypothesis testing problem. While these bounds where available in the literature, they have not found widespread application in communication theory. We give a simple binary hypothesis testing problem where such bounds will be very helpful

This work was supported in part by the National Science Foundation, and by the California Institute of Telecommunications and Information Technology at the University of California San Diego. A version of this paper will be presented at the 44-th Allerton Conference on Communication, Control, and Computing (Allerton'06).

in analyzing the optimal decision rule. We then derive a new upper bound on probability of error in multi-hypothesis testing of the form

$$P_e \leqslant 1 - 2^{-H(X|Y)} \quad (2)$$

which like the equivocation bound [R66, HR70] relates $P_e$ to the conditional entropy. But unlike the classical equivocation bounds, the new bound is always bounded below 1 and never gets too loose to be uninformative.

Next we use these bounds and a random coding [Gal65, SGB67] argument to obtain a sphere packing lower bound on probability of error under MAP of the ensemble of random codes for any channel in a subsequent section. Then we specialize it to the case of a memoryless channel to obtain a lower bound on equivocation for most random codes,

$$H(X|Y) \geqslant \frac{N}{2}(R - \rho) \quad (3)$$

where $N$ is the block length of a rate $R$ code and $\rho$ is a function of the ápriori input probability distribution and the channel likelihood function. For a discrete memoryless channel,

$$\rho_{p_x(.)} \stackrel{\text{def}}{=} 2\log_2\left(\sum_{j \in \mathcal{J}} \sqrt{\sum_{k \in \mathcal{K}} p_x(k) p_{y|x}(j|k)^2}\right)$$

where the DMC channel transition function given by $p_{y|x}(.)$, while $p_x(.)$ is some probability distribution on the input alphabet and $\mathcal{K}$ and $\mathcal{J}$ are the input and output alphabets respectively. This also leads us to an upper bound on the mutual information and hence the capacity of such channels. For a discrete memoryless channel $C \leqslant \max_{p_x(.)} \{\rho_{p_x(.)}\}$.

In the next section, we derive some tight bounds on the probability of error under MAP. Some of these bounds are well known [Vaj68, Tou72, Dev74].

## II. Bounds on Error Probability under MAP Criterion

Consider a $M$-ary hypothesis testing problem. Let our $M$ hypotheses be denoted as $\{h_i : i \in \{1, 2, \ldots, M\}\}$ and their corresponding ápriori probabilities be given by $\{\pi_i : i \in \{1, 2, \ldots, M\}\}$. Also let the noisy observation be $y$. For MAP decision decoding, the conditional probability of error is,

$$P_{e|y} = 1 - \max_{i \in \{1,2,\ldots,L\}} P(h_i|y)$$

while $\sum_i P(h_i|y) = 1$.

### A. Bounds on Probability of error for binary hypothesis testing

We begin by looking at the binary hypothesis problem. If we use the MAP criterion, the average probability of error is given by [HR70],

$$P_e = E_y[1 - \max_{i \in \{1,2\}} P(h_i|y)] = E_y[\min_{i \in \{1,2\}} P(h_i|y)]$$

for the two hypothesis case. By an application of the well known weighted geometric mean inequality, we immediately obtain the upper bound:

$$\begin{aligned} P_e &= E_y[\min_{i \in \{1,2\}} P(h_i|y)] \\ &\leqslant \min_{0 \leqslant \alpha \leqslant 1} E_y[P(h_1|y)^\alpha P(h_2|y)^{(1-\alpha)}] \end{aligned}$$

which is the popular Chernoff bound [Che52]. For the special case of $\alpha = 1/2$, this reduces to the Bhattacharyaa bound [Kai67]. The Chernoff bound is not particularly convenient to use due to the required optimization outside the expectation, while the Bhattacharyaa bound is very loose.

Using the negative power mean inequalities, we can do much better. We have for any $\beta < 0$,

$$\begin{aligned} P_e &= E_y[\min_{i \in \{1,2\}} P(h_i|y)] \\ &\leqslant 2^{-1/\beta} E_y[(P(h_1|y)^\beta + P(h_2|y)^\beta)^{1/\beta}] \end{aligned}$$

While the bound gets tighter as $\beta \to -\infty$, for most practical purposes, we can limit to the case $\beta = -1$, which corresponds to the harmonic mean. After simplifications, we have,

$$\begin{aligned} HM(P(h_1|y), P(h_2|y)) &= 2P(h_1|y)P(h_2|y) \\ &= 1 - P(h_1|y)^2 - P(h_2|y)^2 \end{aligned}$$

where $HM$ denotes the harmonic mean. So, we have the following pair of upper and lower bounds on the conditional probability of error, $P_{e|y}$:

$$P(h_1|y)P(h_2|y) \leqslant P_{e|y} \leqslant 2P(h_1|y)P(h_2|y)$$

and for $P_e$:

$$P_{LB} \leqslant P_e \leqslant 2P_{LB} \quad (4)$$

where $P_{LB} \stackrel{\text{def}}{=} \int_y \frac{\Pr(h_1)\Pr(h_2)P(y|h_1)P(y|h_2)}{\Pr(h_1)P(y|h_1)+\Pr(h_2)P(y|h_2)} dy$. It should be noted that we also obtained a convenient lower-bound on $P(e)$, which is one half the upper-bound, by making use of the properties of harmonic means. We will refer to this pair as the **harmonic bound**. The factor of 2 guarantee in tightness between upper and lower bounds in the probability of error is usually enough for most practical applications. Given in Appendix I is an example of a binary hypothesis testing problem where the exact performance of the optimal decision rule is difficult to determine and bounds are useful.

One may ask if there are $M$-ary extensions to the harmonic bound. It turns out that this is indeed the case. Though motivated due to other reasons, such bounds are well known in the literature [Vaj68, Tou72, Dev74], with suggested applications in multi-hypothesis pattern recognition. We look at some of these extensions in the next two sections. We will also derive a new inequality and upper bound during the process.

## B. Some Inequalities for bounded positive sequences

In this section we first consider a few well known inequalities for bounded positive valued sequences. We then derive a new (to the authors) inequality. In the rest of the section, $\{a_i : i \in \{1, 2, \ldots, M\}\}$ is assumed to be a discrete probability distribution. $M$ is either finite or countably infinite.

We will need some well known inequalities [BB61, Vaj68, Tou72, Dev74] for proving our main results. For the sake of completeness, we give a proof in the appendix.

**Lemma II.1.**

(i) $\max_i \{a_i\} \leqslant \sqrt{\sum_i a_i^2}$

(ii) $\max_i \{a_i\} \geqslant \sum_i a_i^2$

(iii) $2(1 - \sqrt{\sum_i a_i^2}) \geqslant (1 - \sum_i a_i^2)$

*Proof.* Please see Appendix II. ∎

The following inequality is new to the authors. Motivated by continuity considerations, the convention $0 \log_2(0) = 0$ and $0^0 = 1$ is adopted.

**Lemma II.2.** $\sum_i a_i^2 \geqslant 2^{-H(\underline{a})} = \prod_i a_i^{a_i}$

*Proof.* We use induction.

(1) $\underline{M = 1}$: $a_1 = 1$ is the only possibility and claim holds.
(2) $\underline{M = m+1}$: We prove the $M = (m+1)$ case assuming that the claim is true for $M = m$. Consider the normalized sequence, $a'_i = \frac{a_i}{\sum_{i=1}^m a_i} = \frac{a_i}{1 - a_{m+1}}$. One may take $a_{m+1} \neq 1$, for otherwise, the claim is trivially true. By induction hypothesis,

$$\sum_{i=1}^m (a'_i)^2 \geqslant \prod_i (a'_i)^{a'_i}$$

After some algebra, we get

$$\sum_{i=1}^m a_i^2 \geqslant (1 - a_{m+1}) \left( \prod_i a_i^{a_i} \right)^{\frac{1}{(1-a_{m+1})}}$$

We are done if we show that $x^2 + (1-x)y^{\frac{1}{(1-x)}} \geqslant x^x y$ when $0 \leqslant x, y < 1$. To see that this is true, let us fix $0 \leqslant x = \alpha < 1$ and consider the function $f(y) \triangleq \alpha^2 + (1-\alpha)y^{\frac{1}{(1-\alpha)}} - \alpha^\alpha y$.
Taking derivatives, $f'(y) = y^{\frac{1}{(1-\alpha)} - 1} - \alpha^\alpha$ and $f''(y) = \left( \frac{1}{(1-\alpha)} - 1 \right) \cdot y^{\frac{1}{(1-\alpha)} - 2} \geqslant 0$ because $0 \leqslant \alpha < 1$. So $f(y)$ is a convex $\cup$ function of $y$ and has a global minimum of 0 at $y = \alpha^{1-\alpha}$.

This completes the proof. ∎

## C. Tight Bounds on probability of error in multi-hypothesis testing

One can substitute $P(\boldsymbol{h}_i|\boldsymbol{y})$ for $a_i$ in the inequalities derived in the previous section. Then we have the following:

$$1 - \sqrt{\sum_i P(\boldsymbol{h}_i|\boldsymbol{y})^2} \leqslant P_{e|\boldsymbol{y}} \leqslant 2 - 2\sqrt{\sum_i P(\boldsymbol{h}_i|\boldsymbol{y})^2} \quad (5)$$

A related pair of bounds

$$\tfrac{1}{2} - \tfrac{1}{2}\sum_i P(\boldsymbol{h}_i|\boldsymbol{y})^2 \leqslant P_{e|\boldsymbol{y}} \leqslant 1 - \sum_i P(\boldsymbol{h}_i|\boldsymbol{y})^2 \quad (6)$$

was first discussed in [Vaj68] in the context of *Vajda's quadratic entropy* and later by Toussaint [Tou72] who proposed the *quadratic mutual information* and by Devijver [Dev74], who popularized a closely connected measure called the *Bayesian Distance* in pattern recognition. Devijver also mentions the lower bound in (5). The later pair (6) can be thought of as an $M$-ary extension to the harmonic mean bound.

## D. An Improvement over Rényi's Equivocation Bound

Now we consider upper bounds relating $P_e$ with the equivocation. In [R66], Rényi derived the bound:

$$P_{e|\boldsymbol{y}} \leqslant H\left(P(\boldsymbol{h}|\boldsymbol{y})\right) \quad (7)$$

Hellman and Raviv later improved this bound in [HR70] to:

$$P_{e|\boldsymbol{y}} \leqslant \tfrac{1}{2} H\left(P(\boldsymbol{h}|\boldsymbol{y})\right) \quad (8)$$

These relations are not bounded and can get very loose when there are many hypotheses with roughly equal áposteriori probabilities. Using the new inequality from Lemma II.2 we get:

$$P_{e|\boldsymbol{y}} \leqslant 1 - \sum_i P(\boldsymbol{h}_i|\boldsymbol{y})^2 \leqslant 1 - 2^{-H(P(\boldsymbol{h}|\boldsymbol{y}))} \quad (9)$$

where $H(.)$ denotes the usual entropy function. Recalling that $H(X|Y) \stackrel{\text{def}}{=} E_y \left[ H\left(P(\boldsymbol{h}|\boldsymbol{y})\right) \right]$,

$$P_e \leqslant 1 - E_y \left[ 2^{-H(P(\boldsymbol{h}|\boldsymbol{y}))} \right] \leqslant 1 - 2^{-H(X|Y)} \quad (10)$$

where we used the fact that $2^{-z}$ is a convex $\cup$ function of $z$ and the Jensen's inequality. This is a new bound which relates equivocation to the Bayes risk. It is also an improvement over the Rényi and Hellman-Raviv bounds. Expanding the bound in (10) as a power series,

$$P_e \leqslant 1 - 2^{-H(X|Y)} = \sum_{n=1}^{\infty} (-1)^{n+1} \frac{(H(X|Y) \ln 2)^n}{n!} \quad (11)$$

which is always better than the Rényi bound and at most a factor of $\ln 4 < 1.4$ worse than the Hellman-Raviv equivocation bound – this is quite acceptable for most purposes. While for the binary hypothesis case the new bound of (10) is not as tight as the equivocation bound of (8), as the number of hypothesis increases the equivocation can far exceed 1. This makes both the Rényi and Hellman-Raviv bounds very loose. For example when $P(\boldsymbol{h}_i|\boldsymbol{y}) = \frac{1}{M}$, the Hellman-Raviv equivocation bound is not informative at a loose $\log_2 \sqrt{M}$, while the new bound gives a tight $1 - 2^{-\log_2 M} = \frac{M-1}{M}$.

Comparing the various bounds, the Bayesian distance based bounds of (5) and (6) are far tighter than both the conditional entropy based bounds (10), (8) and the well known union bound using only pairwise error event probabilities. In Figure 1, we can see the various bounds discussed above for the binary hypothesis case.

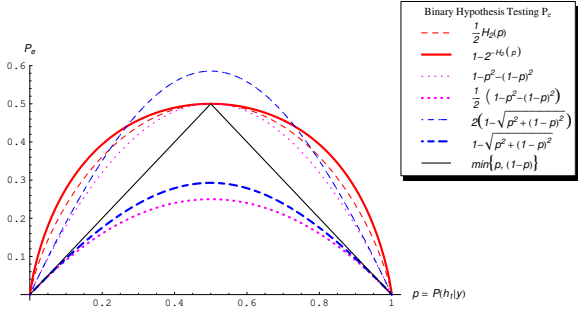

**Figure 1:** Probability of error $P_e$ and various bounds on it for binary hypothesis testing.

There are many instances of $M$-ary hypothesis testing in communication theory where the bounds discussed in this section can be valuable fundamental analysis tools. The rest of the paper uses only the bounds given by (5) and (10).

### III. A RANDOM CODING SPHERE PACKING LOWER BOUND ON $\overline{P}_e$ AND EQUIVOCATION

In this section, we wish to apply the random coding argument [Gal65, SGB67], to obtain a lower bounds on the ensemble average of expected probability of error under MAP decoding for any channel.

We have,

$$\overline{P}_e = E_{\boldsymbol{y}}[P_{e|\boldsymbol{y}}] \geqslant E_{\boldsymbol{y}}\left[1 - \sqrt{\sum_i P(\boldsymbol{h}_i|\boldsymbol{y})^2}\right]$$

Now consider the ensemble of random codes. Each codeword in a random code in this ensemble is chosen independently and at random from the set of all possibilities with a probability of $P(\boldsymbol{x})$. We will use the overbar to denote the ensemble average. The following is immediately obtained

$$\overline{P}_e \geqslant 1 - \overline{E_{\boldsymbol{y}}\left[\sqrt{\sum_i P(\boldsymbol{h}_i|\boldsymbol{y})^2}\right]} \quad (12)$$

where we made use of the linearity property of expectation. There is also a corresponding upper bound:

$$\overline{P}_e \leqslant 2 - 2\overline{E_{\boldsymbol{y}}\left[\sqrt{\sum_i P(\boldsymbol{h}_i|\boldsymbol{y})^2}\right]}$$

In this paper, we will not be further concerned with the above upper bound on $\overline{P}_e$. Instead we concentrate on inequality (12).

The inequality in (12) can be further simplified when expanded out in terms of the input and output probability distributions and the channel likelihood function as follows:

$$\overline{P}_e \geqslant 1 - \overline{E_{\boldsymbol{y}}\left[\sqrt{\sum_i P(\boldsymbol{h}_i|\boldsymbol{y})^2}\right]}$$

$$= 1 - \overline{\sum_{\boldsymbol{y}} P(\boldsymbol{y}) \cdot \sqrt{\sum_i P(\boldsymbol{h}_i|\boldsymbol{y})^2}}$$

$$= 1 - \overline{\sum_{\boldsymbol{y}} P(\boldsymbol{y}) \cdot \sqrt{\sum_i \left(\frac{P(\boldsymbol{h}_i) \cdot P(\boldsymbol{y}|\boldsymbol{h}_i)}{P(\boldsymbol{y})}\right)^2}}$$

$$= 1 - \overline{\sum_{\boldsymbol{y}} \sqrt{\sum_i P(\boldsymbol{h}_i)^2 \cdot P(\boldsymbol{y}|\boldsymbol{h}_i)^2}}$$

$$= 1 - \sum_{\boldsymbol{y}} \overline{\sqrt{\sum_i P(\boldsymbol{h}_i)^2 \cdot P(\boldsymbol{y}|\boldsymbol{h}_i)^2}} \quad (13)$$

where we used linearity of expectation in the last step.

Due to the tightness of the bound on $P_{e|\boldsymbol{y}}$ which we used initially, the ensemble average lower bound of (13) is also tight within a factor of 2. However, the expression is not easily amenable to further simplification. We now apply Jensen's inequality to obtain a looser yet considerably simpler lower bound:

$$\overline{P}_e \geqslant 1 - \sum_{\boldsymbol{y}} \overline{\sqrt{\sum_i P(\boldsymbol{h}_i)^2 \cdot P(\boldsymbol{y}|\boldsymbol{h}_i)^2}}$$

$$\geqslant 1 - \sum_{\boldsymbol{y}} \sqrt{\sum_i P(\boldsymbol{h}_i)^2 \cdot \overline{P(\boldsymbol{y}|\boldsymbol{h}_i)^2}} \quad (14)$$

Here we used the fact that $\sqrt{x}$ is a concave $\cap$ function of $x$. Then by Jensen's inequality, $E_x[\sqrt{f(x)}] \leqslant \sqrt{E_x[f(x)]}$.

Let us also assume without loss of generality that our hypothesis (codeword) $\boldsymbol{h}_i$ occurs with an ápriori probability $\pi_i$. In particular for the equiprobable case, $\pi_i = \frac{1}{M}$, where $M$ is the total number of codewords in the code under consideration. We get,

$$\overline{P}_e \geqslant 1 - \sum_{\boldsymbol{y}} \sqrt{\sum_i \pi_i^2 \cdot \overline{P(\boldsymbol{y}|\boldsymbol{h}_i)^2}}$$

$$= 1 - \sum_{\boldsymbol{y}} \sqrt{\sum_i \pi_i^2 \cdot \overline{P(\boldsymbol{y}|\boldsymbol{h}_i)^2}}$$

$$= 1 - \sqrt{\sum_i \pi_i^2} \cdot \sum_{\boldsymbol{y}} \sqrt{\sum_{\boldsymbol{x}} P(\boldsymbol{x}) P(\boldsymbol{y}|\boldsymbol{x})^2} \quad (15)$$

as the ensemble average is independent of the particular hypothesis (transmitted codeword). In the above equation, $\boldsymbol{x}$ is a random vector drawn from the ensemble according to a probability distribution $P(\boldsymbol{x})$.

Ideally, we would like to optimize on the codeword

ápriori probabilities subject to certain constraints:

$$\boxed{\begin{aligned} \text{Minimize} \quad & -\sum_i \pi_i^2 \quad \text{subject to,} \\ -\sum_i \pi_i \log_2 \pi_i &= NR \\ \sum_i \pi_i &= 1 \quad \text{and} \\ \pi_i &\geqslant 0, \forall i \end{aligned}} \quad (16)$$

where, $N$ is the block-length of the code and $R$ is its *information rate* in (bits/use). If we set $NR = \log_2 M$, the only feasible solution is $\pi_i = \frac{1}{M}$. This choice of ápriori is also justified by the Channel Coding Theorem for DMC, where an equally likely selection of codewords is shown to achieve channel capacity for an ensemble of random codes. With this setting, we get:

$$\overline{P}_e \geqslant 1 - \frac{1}{\sqrt{M}} \cdot \sum_{\mathbf{y}} \sqrt{\sum_{\mathbf{x}} P(\mathbf{x}) P(\mathbf{y}|\mathbf{x})^2} \quad (17)$$

We now specialize (13) to the case of a discrete memoryless channel. Recall that, for a discrete memoryless channel which is discrete in time,

$$P(\mathbf{y}|\mathbf{x}) = \prod_n p_{y|x}(y_n|x_n)$$

By the proof of the Channel Coding Theorem [Sha48], we know that for random ensembles of codes where codewords are chosen such that each symbol is chosen independently of each other using a probability distribution given by $p_x(.)$, the ensemble average probability of decoding (under the suboptimal jointly typical decoding) tends to zero as block-lengths tend to infinity. We will also likewise specialize to such an ensemble of codes, without any loss of generality. For this special class of codes, $P(\mathbf{x}) = \prod_n p_x(x_n)$. So,

$$\begin{aligned} \overline{P}_e &\geqslant 1 - \frac{1}{\sqrt{M}} \cdot \sum_{\mathbf{y}} \sqrt{\sum_{\mathbf{x}} \prod_n p_x(x_n) p_{y|x}(y_n|x_n)^2} \\ &= 1 - \frac{1}{\sqrt{M}} \cdot \sum_{\mathbf{y}} \sqrt{\prod_n \sum_{x_n \in \mathcal{K}} p(x_n) p(y_n|x_n)^2} \\ &= 1 - \frac{1}{\sqrt{M}} \cdot \prod_n \sum_{y_n \in \mathcal{J}} \sqrt{\sum_{x_n \in \mathcal{K}} p(x_n) p(y_n|x_n)^2} \\ &= 1 - \frac{1}{\sqrt{M}} \left( \sum_{j \in \mathcal{J}} \sqrt{\sum_{k \in \mathcal{K}} p_x(k) p_{y|x}(j|k)^2} \right)^N \end{aligned} \quad (18)$$

where $\mathcal{K}$ and $\mathcal{J}$ are the input and output alphabets respectively. In performing the above simplifications, we made repeated use of interchanging summation and product.

Let us define a parameter $\rho$ as follows:

$$\rho \triangleq 2 \log_2 \left( \sum_{j \in \mathcal{J}} \sqrt{\sum_{k \in \mathcal{K}} p_x(k) p_{y|x}(j|k)^2} \right) \quad (19)$$

### A. Continuous Alphabet channels

It is usual to define [McE02] a continuous alphabet channel to be memoryless when for any finite quantization of input and output alphabet, the quantized discrete channel is memoryless. Under this definition and if we assume that the associated probability measures are regular [Fel70], then the corresponding result holds for any memoryless channel, where the summations are replaced by appropriate Riemann integrations. So for well behaved continuous alphabet memoryless channels,

$$\overline{P}_e \geqslant 1 - \frac{1}{\sqrt{M}} \left( \int_{\beta \in \mathcal{J}} \sqrt{\int_{\alpha \in \mathcal{K}} p(\alpha) p(\beta|\alpha)^2 \, d\alpha} \, d\beta \right)^N \quad (20)$$

Here we define $\rho$ as follows:

$$\rho \triangleq 2 \log_2 \left( \int_{\beta \in \mathcal{J}} \sqrt{\int_{\alpha \in \mathcal{K}} p_x(\alpha) p_{y|x}(\beta|\alpha)^2 \, d\alpha} \, d\beta \right) \quad (21)$$

### B. A Lower Bound on Equivocation

Earlier we chose $M = 2^{NR}$. Thus for either a discrete alphabet or a well behaved continuous alphabet memoryless channel,

$$\overline{P}_e \geqslant 1 - 2^{-\frac{N}{2}(R-\rho)} \quad (22)$$

Using Jensen's inequality and (10) we get:

$$\overline{P}_e \leqslant 1 - 2^{-\overline{H(X|Y)}} \quad (23)$$

On combining (22) and (23) we have proved:

**Theorem III.1** *Most codes in the ensemble of capacity achieving random codes considered in this section when used over a memoryless channel satisfy the lower bound on equivocation:*

$$H(X|Y) \geqslant \frac{N}{2}(R - \rho) \quad (24)$$

Another application of (22) is in upper bounding the capacity of memoryless channels. In Appendix III this is explored further. Several simple examples are also provided.

---

APPENDIX I
A BINARY HYPOTHESIS TESTING PROBLEM

**Example I.1** Consider the following binary hypothesis testing problem:

$$\begin{aligned} h_1 &: \ y = n_1 \\ h_2 &: \ y = n_2 \end{aligned}$$

where $n_1$ is distributed with a pdf given by

$$P(y|h_1) = f_{n_1}(z) = \tfrac{2}{3}(\cos(t/2))^2 e^{-|t|}$$

having unit variance and $n_2$ has a Gaussian pdf given by

$$P(y|h_2) = f_{n_2}(z) = \sqrt{\frac{\gamma}{\pi}} e^{-\gamma z^2}$$

and the ápriori probabilities are $\Pr(h_i) = \frac{1}{2}$.

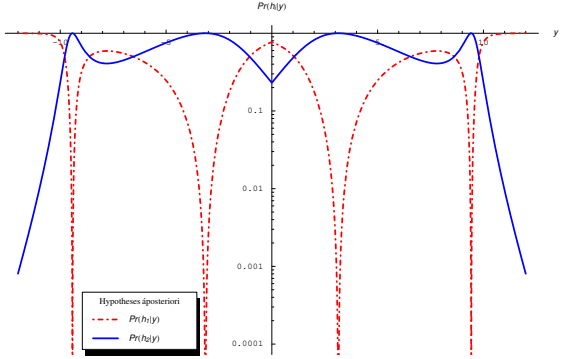

**Figure 2:** The áposteriori probabilities corresponding to the two hypothesis when $\gamma = \frac{1}{8}$. The decision region boundaries are marked by the crossings of the two plots.

From Figure 2, we can see that the optimum decision region for this problem is very difficult to compute in general. As a result the exact Bayes risk is also difficult to obtain, and tight bounds on $P_e$ are of interest. There are no tightness guarantees for either the Bhattacharyaa or Chernoff bound, while the harmonic bound of (4) is very tight as can be observed in Figure 3.

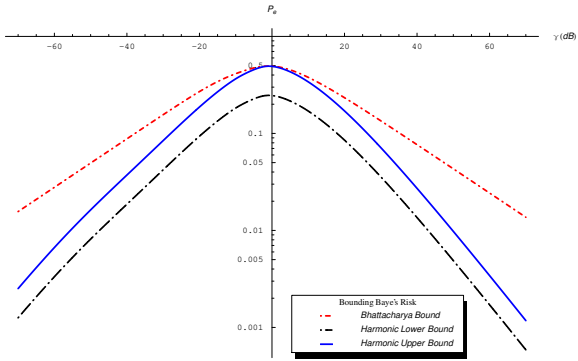

**Figure 3:** Bounds on Bayes risk in the two hypothesis testing problem.

## APPENDIX II
## PROOF OF LEMMA II.1

(i) Our proof is by mathematical induction.
   (1) $\underline{M = 1}$: $a_1 = 1$ is the only possibility, and claim is obvious.
   (2) $\underline{M = m + 1}$: Let us hypothesize that the claim is true for $M = m$. We now prove the $M = (m+1)$ case. Let us use the notation, $\mu_\ell(\underline{a}) \triangleq \max_{i=1}^{\ell} \{a_i\}$.

$$\mu_{m+1}(\underline{a}) = \max\{\mu_m(\underline{a}), a_{m+1}\}$$

Consider the normalized sequence, $a'_i = \frac{a_i}{\sum_{i=1}^{m} a_i} = \frac{a_i}{1 - a_{m+1}}$. We can safely take $a_{m+1} \neq 1$, for otherwise, the claim is trivially true. By induction hypothesis,

$$\mu_m(\underline{a'}) \leqslant \sqrt{\sum_{i=1}^{m}(a'_i)^2}$$

This gives us $\mu_m(\underline{a}) \leqslant \sqrt{\sum_{i=1}^{m} a_i^2}$. So we are done if we prove that,

$$\max\left\{\sqrt{\sum_{i=1}^{m} a_i^2},\ a_{m+1}\right\} \leqslant \sqrt{\sum_{i=1}^{m+1} a_i^2}$$

But we know that $x, y \leqslant \max\{x, y\} \leqslant \sqrt{x^2 + y^2}$ by considering each case separately.

(ii) Clearly, $\max_i \{a_i\} = \max_i \{a_i\} \cdot \sum_i a_i \geqslant \sum_i a_i^2$.
(iii) We need to only observe that $2(1 - x) \geqslant 1 - x^2$. ∎

## APPENDIX III
## AN UPPER BOUND ON MUTUAL INFORMATION AND CAPACITY

By observing the bound of (22), we see that the bound is trivial whenever $R \leqslant \rho$. However, when $R > \rho$, $\overline{P}_e \to 1$ exponentially. On the other hand, for the ensemble of codes we considered, the Channel Coding Theorem says that the ensemble probability of error can be made arbitrarily small, using even the suboptimal jointly typical decoding algorithm at the decoder whenever rate $R$ is below the mutual information between channel input and output. Therefore we have proved the following upper bound on mutual information (and hence the capacity) of a memoryless channel:

$$I(X;Y) \leqslant \rho_{p_x(.)} \quad (25)$$
$$C = \max_{p_x(.)}\{I(X;Y)\} \leqslant \max_{p_x(.)}\{\rho_{p_x(.)}\} \quad (26)$$

where $\rho$ is given by either (19) or (21).

### A. Discussion and Some Examples

The practical usefulness for the derived upper bound depends on two factors, namely the tightness of the bound and the ease of computation. In the derivation of the upper-bound for mutual information, the only loss in tightness is in the use of Jensen's inequality during the ensemble averaging process. The function $\rho$ has to be maximized over all possible input distributions $p_x(.)$ to obtain the upper bound. The required optimization can make the computation of the bound difficult. However, the expression is considerably simpler than the expression for mutual information and may be easier to deal with for some particular channel.

Below, results are presented for some very common channels. The tightness of the upper bound on capacity is found to be acceptable.

*1) Binary Symmetric Channels:* For a BSC with crossover probability $p$, using the capacity achieving input distribution we get, $\rho = 1 + \log_2\left(p^2 + (1-p)^2\right)$ and the capacity is well known to be $C = 1 - H_2(p)$, where $H_2(.)$ is the binary entropy function. See Figure 4(a).

*2) Binary Erasure Channels:* For a BEC with probability of erasure $\epsilon$, again using the capacity achieving input distribution we get $\rho = 2\log_2\left(\sqrt{2} - (\sqrt{2}-1)\epsilon\right)$ whereas, the capacity is given by $C = 1 - \epsilon$. Both are shown in Figure 4(b).

*3) Binary Input – zero-mean AWGN – Soft Output Channel:* For a memoryless channel with binary input ($\pm 1$) and soft output and affected by additive white Gaussian noise of zero-mean, using the capacity achieving input distribution of $[\frac{1}{2}, \frac{1}{2}]$ probability, we get,

$$\rho = -\log_2 4\pi\sigma^2 + \\ 2\log_2\left(\int_{y=-\infty}^{\infty} \sqrt{e^{-\frac{(y-1)^2}{\sigma^2}} + e^{-\frac{(y+1)^2}{\sigma^2}}}\, dy\right)$$

and the capacity is given by:

$$C = -\frac{1}{2}\log_2\left(2\pi e\sigma^2\right) \\ -\int_{y=-\infty}^{\infty}\left(\frac{e^{-\frac{(y-1)^2}{2\sigma^2}}}{2\sqrt{2\pi\sigma^2}} + \frac{e^{-\frac{(y+1)^2}{2\sigma^2}}}{2\sqrt{2\pi\sigma^2}}\right)\cdot \\ \log_2\left(\frac{e^{-\frac{(y-1)^2}{2\sigma^2}}}{2\sqrt{2\pi\sigma^2}} + \frac{e^{-\frac{(y+1)^2}{2\sigma^2}}}{2\sqrt{2\pi\sigma^2}}\right)dy$$

where, the noise is distributed as $\mathcal{N}(0,\sigma^2)$ with variance $\sigma^2 = \frac{N_0}{2}$. In this case, the numerical integration required for computing the capacity is unstable at very low $E_b/N_0$, due to the presence of the $\log(\cdot)$ function in the integrand. However, the upper bound integration remains stable up to a much lower $E_b/N_0$. See Figure 4(c).

The definition of capacity or mutual information was not needed in the derivation of the capacity bound in this section because of the use of the random coding argument. It was pointed out by Prof. Shlomo Shamai (Shitz) that it is possible to derive the above bound using only the functional definition of mutual information (1) and Jensen's inequality. For most practical applications, a tightness guarantee is also desirable.

*Acknowledgments:* The authors wish to thank Prof. Shlomo Shamai (Shitz) for pointing out a simple alternate derivation of the capacity bound.

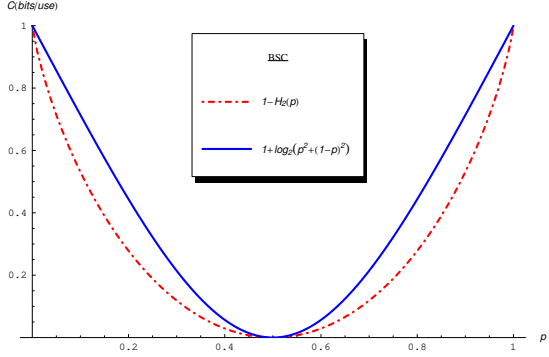

(a) BSC with crossover probability of $p$.

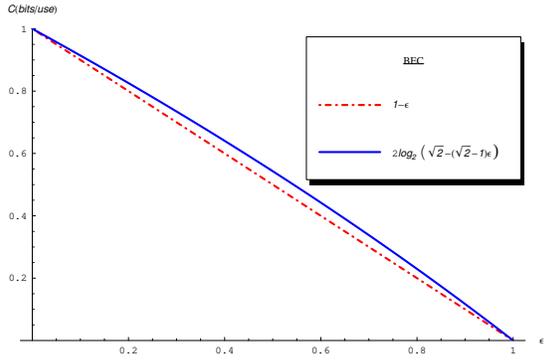

(b) BEC with probability of erasure $\epsilon$.

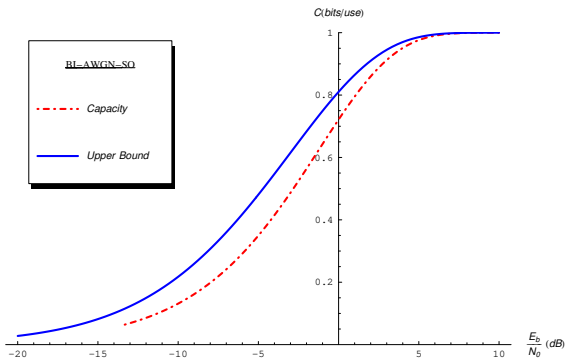

(c) Binary input, AWGN, soft output channel. The binary inputs are ($\pm 1$) and AWGN has the distribution, $\mathcal{N}(0, \frac{N_0}{2})$

**Figure 4:** Actual capacity and upper bounds on the capacity for some common channels.